\documentclass[10pt,journal,twocolumn,twoside]{IEEEtran} 
\normalsize
\usepackage{fancyhdr,lipsum}
\usepackage{filecontents}
\usepackage[dvips]{color, graphicx}
\usepackage[cmex10]{amsmath}
\usepackage{array}
\usepackage{booktabs}
\usepackage{times}
\usepackage{flushend}
\usepackage{amsthm}
\usepackage{amsmath,amssymb}
\usepackage{slashbox,multirow}
\usepackage[square, comma, sort&compress, numbers]{natbib}
\ifCLASSOPTIONcompsoc
\usepackage[caption=false,font=normalsize,labelfon
t=sf,textfont=sf]{subfig}
\else
\usepackage[caption=false,font=footnotesize]{subfi
g}
\fi
\makeatletter
\newif\if@restonecol
\makeatother

\usepackage[ruled,vlined,linesnumbered]{algorithm2e}

\ifCLASSINFOpdf
\else
\fi
\begin{document}
\title{A Correlation-Breaking Interleaving of \\
Polar Codes in Concatenated Systems}
\author{Ya~Meng,~\IEEEmembership{~Student Member,~IEEE,}
Liping~Li,~\IEEEmembership{~Member,~IEEE},
Chuan~Zhang,~\IEEEmembership{~Member,~IEEE}
\thanks{This work was supported in part by National Natural Science Foundation of China through grant 61501002, in part by Natural Science Project of Ministry of Education of Anhui through grant KJ2015A102, in part by Talents Recruitment Program of Anhui University, in part by the Key Laboratory Project of the Key Laboratory of Intelligent Computing and Signal Processing of the Ministry of Education of China, Anhui University. This
paper was presented in part at the IEEE Vehicular Technology Conference Fall, Montreal, 2016.
}
\thanks{Ya Meng and Liping Li are with the Key Laboratory of Intelligent Computing and Signal Processing of the Ministry of Education of China, Anhui University. Chuan Zhang is with the National Mobile
Communications Research Laboratory, Southeast University, Nanjing, China
(e-mail: mengya@ahu.edu.cn; liping\_li@ahu.edu.cn; chzhang@seu.edu.cn). (Corresponding author: Liping Li.)}
}
\markboth{IEEE Transactions on Communications}%
{Submitted paper}

\maketitle
\begin{abstract}
It is known that the bit errors of polar codes with successive cancellation (SC) decoding are coupled. However, existing concatenation schemes of polar codes with other error correction codes rarely take this coupling effect into consideration.
To achieve a better  error performance of concatenated systems with polar codes as inner codes,
one can divide all bits in an outer block into different polar blocks to completely de-correlate the possible coupled errors in the transmitter side.
We call this interleaving a
blind interleaving (BI) which serves as a benchmark. Two BI schemes, termed BI-DP and BI-CDP, are proposed
in the paper. To better balance performance, memory size, and the decoding delay from the de-interleaving, a novel interleaving scheme, named the correlation-breaking interleaving (CBI), is proposed. The CBI breaks the correlated information bits based on the error correlation pattern proposed and proven in this paper.
The proposed CBI scheme is general in the sense that any error correction code can serve as the outer code.
In this paper, Low-Density
Parity-Check (LDPC) codes and BCH codes
are used as two examples of the outer codes of the interleaving scheme.
The  CBI scheme 1) can keep the simple SC polar decoding while achieving a better error performance than the state-of-the-art (SOA) direct concatenation of polar codes with LDPC codes and BCH codes; 2) achieves a comparable error performance as the BI-DP scheme with a smaller memory size and a shorter decoding delay.
Numerical results are provided to verify the performance of the BI schemes and the CBI scheme.
\end{abstract}

\section{Introduction}\label{sec_Intro}
\IEEEPARstart{T}{he} channel polarization and polar codes were discovered by Ar\i kan in \cite{arikan_iti09} which made a great progress in coding theory. Polar codes provably achieve the capacity of symmetric binary-input discrete memoryless channels (B-DMCs) with a low encoding and decoding complexity. The encoding and decoding process (with successive cancellation, SC) can be implemented with a complexity of $\mathcal{O}(N \log N)$, where $N$ is the block length.
The idea of polar codes is to transmit information bits on noiseless bit channels while fixing the information bits on
the completely noisy bit channels.
The fixed bits (also called as the frozen bits) are made known to both the transmitter and receiver. The standard format of polar codes in \cite{arikan_iti09} is non-systematic. Later,
the systematic version of polar codes was proposed in \cite{arikan_icl11}.
The construction of polar codes is studied in \cite{vardy_it13,mori_icl09,trifonov_itc12,wu_icl14}
and the hardware implementation is presented in \cite{C.Z_itsp13,vardy_jsac14,C.Z_itcs14}.

To improve the polar code performance with the finite block length, various decoding
processes \cite{arikan_icl08,urbanke_isit09,vardy_it15,niu_itc13}
and concatenation schemes \cite{ eslami_isit11,guo_isit14,effros_isit10, jiang_cl16,song_el17}
were proposed.
The decoding processes in these works have higher complexity than the original SC decoding of \cite{arikan_iti09}.
The performance improvements in these decoding algorithms are at the cost of the decoding complexity.
The introduction of the systematic polar codes \cite{arikan_icl11} provides a new way to improve
bit error rate (BER) performance
 while still maintaining almost the same decoding complexity as non-systematic polar codes.

The spreading effect of the error bit on the following decoding steps results in  the
known error propagation problem.
The better BER performance of systematic polar codes can be thought of coming from the error-decoupling. The non-systematic encoding is $x_1^N=u_1^NG$, where the vector $u_1^N$ contains the source bits and $G$ is the generator matrix. From the two-step decoding of systematic polar codes (first estimating $\hat{u}_1^N$ and then calculating $\hat{x}_1^N$ from it),
this decoupling must be accomplished through the re-encoding
$\hat{x}_1^N=\hat{u}_1^NG$ after obtaining the estimate $\hat{u}_1^N$.
From $\hat{x}_1^N=\hat{u}_1^NG$ and
that the number of errors in $\hat{x}_1^N$ is smaller than that of $\hat{u}_1^N$, it can be concluded that the
coupling of the errors in $\hat{u}_1^N$ is controlled by the columns of $G$.
A proposition of this error correlation pattern is formally stated and proven in this paper.

Two blind interleaving (BI) schemes are presented to de-correlate the coupled errors.
A concatenation scheme, which divides all bits in an outer code block into different polar blocks to completely de-correlate
the possible coupled errors, is first introduced as a benchmark.
Note that this BI scheme is also called a direct product of the inner
and outer code, termed as BI-DP in the paper.
The BI scheme
can keep the simple SC polar decoding while achieving a better BER performance than the state-of-the-art concatenation of polar codes with outer codes. An improved BI scheme, called `quasi' cyclicly shifted direct product BI (BI-CDP), is introduced to improve the BI-DP scheme. This BI-CDP scheme takes into consideration the different
levels of protection experienced by the information bits in one polar block, and assigns the coded bits from the outer code
into cyclicly shifted information positions of the inner code. This BI-CDP scheme is shown to yield a better
error performance than the BI-DP scheme. Note that the BI-CDP is different from the Twill interleaving in \cite{Stefan_infoTheory02} since it does not require the greatest common divisor (gcd) of the number of the encoder for the inner code and the outer code equal to $1$. In this paper, the number of the encoder for the inner code and the outer code is the code length of the outer code and the number of the information bits of the inner code, respectively.

From the error correlation pattern presented in the paper,
a novel interleaving scheme, named the correlation-breaking interleaving (CBI),
is proposed to better balance among performance,
memory size, and the decoding delay  from the de-interleaving operation.
The proposed CBI scheme divides the information bits into two groups: the group of the correlated bits
$\mathcal{A}_c$ and the group of the uncorrelated bits $\bar{\mathcal{A}}_c$.
Theoretical foundation for procedures to assign elements into these two groups is
provided. As in the BI scheme, the CBI scheme assigns $|\mathcal{A}_c|$ encoded bits from $|\mathcal{A}_c|$ different outer code blocks to the correlated information bits of one polar block. Different from the BI scheme, the CBI scheme assigns $|\bar{\mathcal{A}}_c|$ encoded bits from one
outer code block to
the uncorrelated information bits of one polar block, which
saves the required number of inner polar code blocks. As a result,
the memory size for the de-interleaver and the decoding delay of the outer code can be saved.

Although any outer code works in the CBI scheme, LDPC and BCH codes are chosen in this paper as
examples: the former requiring an iterative soft decoding process while the latter only requiring a simpler syndrome decoder \cite{lin_ecc04}.
Note that the concatenation of polar codes with LDPC codes is studied in
\cite{eslami_isit11} and \cite{guo_isit14} where no interleaving is used and BP
(belief-propagation) decoding is applied for polar codes.
For the ease of description, let us denote polar codes applying SC decoding as POLAR($N$,$K$)-SC,
and polar codes applying BP decoding as POLAR($N$,$K$)-BP, where $K$ is the number of information bits of polar codes in one code block. Also let us denote the direct concatenation
system with a LDPC code as the outer code and a polar code as the inner code as LDPC($N_l$,$K_l$)+POLAR($N$,$K$), where $N_l$ and $K_l$ are the code length and the number of information bits in one LDPC block, respectively.
If a CBI scheme is used between the outer and the inner code, then we denote
such a system as LDPC($N_l$,$K_l$)+CBI+POLAR($N$,$K$).
Similarly, the blind interleaving systems, BI-DP and BI-CDP, are
denoted as LDPC($N_l$,$K_l$)+BI-DP+POLAR($N$,$K$) and LDPC($N_l$,$K_l$)+BI-CDP+POLAR($N$,$K$), respectively.

Simulation results are provided to verify the BER performance of the interleaving schemes in this paper. At a BER $=10^{-4}$, the LDPC($155$,$64$)+CBI+POLAR($256$,$64$)-SC system achieves $1.4$ dB and $1.2$ dB gains over the direct concatenation systems  LDPC($155$,$64$)+POLAR($256$,$64$)-SC and LDPC($155$,$64$)+POLAR($256$,$64$)-BP, respectively.
The LDPC($155$,$64$)+CBI+POLAR($256$,$64$)-SC system also
achieves a comparable performance as that of the LDPC($155$,$64$)+BI-DP+POLAR($256$,$64$)-SC system.
The proposed LDPC($155$,$64$)+BI-CDP+POLAR($256$,$64$)-SC outperforms all the concatenation systems reported. The CBI scheme also works for BCH codes. Here we take the BCH($127$,$57$) with the code length $127$ and the number of information bits $57$ in one code block as an example. At a BER $=10^{-4}$, the BCH($127$,$57$)+CBI+POLAR($256$,$64$)-SC system has a $0.7$ dB gain over the direct concatenation system BCH($127$,$57$)+POLAR($256$,$64$)-SC.

Note that portions of this work are investigated in \cite{ya_vtc16} where the theorems of the error correlation
pattern are not proven and the BI scheme is only one of the two BI schemes in this paper. What's more, the CBI scheme in this paper has a different assignment of the $|\mathcal{A}_c|$ correlated information bits from that in \cite{ya_vtc16}.
In this paper, we
provide the proofs of the theorems, improve the BI scheme and the CBI scheme, and provide examples of the CBI scheme.
Specifically, the contribution of this paper can be summarized as: 1) Theoretically, we prove that the errors
from the SC decoding are coupled. The error correlation pattern is found and proven from two perspectives; 2) Two BI schemes are introduced and a universal CBI scheme (based on the error correlation pattern) is proposed;
3) The CBI scheme is  theoretically explained based on the cyclic arrangements of
coded bits from the outer code to the inner code, and details and examples are provided to illustrate the key parameters.


In this paper,
we use $v_1^N$ to represent a row vector with elements $(v_1,v_2,...,v_N)$.
 For a vector $v_1^N$, the
vector $v_i^j$ is a subvector $(v_i, ..., v_j)$ with $1 \le i,j \le N$. For a given set $\mathcal{A} \in \{1,2,...,N\}$, $v_{\mathcal{A}}$ denotes a subvector with elements in $\{v_i, i \in \mathcal{A}\}$.

The rest of the paper is organized as follows. Section \ref{sec_background} introduces the fundamentals of
non-systematic and systematic polar codes.
The error correlation pattern is raised and proven in section \ref{sec_corre_pattern}.
Section \ref{sec_interleaving} introduces the two BI schemes and proposes the novel CBI scheme.
Section \ref{sec_result} presents the simulation results.
The conclusion remarks are provided at the end.

\section{Background of Polar Codes}\label{sec_background}

In this section, the relevant theories on non-systematic polar codes \cite{arikan_iti09} and systematic polar
codes \cite{arikan_icl11} are presented.

\subsection{Preliminaries of Non-Systematic Polar Codes}\label{sec_background_nonsys}

Let $W:\mathcal{X} \rightarrow \mathcal{Y}$ denote a B-DMC where $\mathcal{X}=\{0,1\}$ is the input and $\mathcal{Y}$ is the output alphabet of the channel. The transition probability is denoted by $W(y|x)$, $x \in \mathcal{X}$, $y \in \mathcal{Y}$.

The generator matrix for polar codes is $G_N= BF^{\otimes n}$ where $B$ is a bit-reversal
matrix, $F=\bigl(\begin{smallmatrix} 1&0 \\ 1&1\end{smallmatrix}\bigr)$, $n=\log_2N$, $N$ is the block length, and
$F^{\otimes n}$ is the $n$th Kronecker power of the matrix $F$ over the binary field $\mathbb{F}_2$. In this paper, we consider an encoding matrix $G_N= F^{\otimes n}$ without the permutation matrix $B$, which only affects the decoding order \cite{arikan_icl11}. For compactness, the subscript of $G_N$ is sometimes omitted as $G$ without causing confusion of the block length $N$.



The channel polarization process is performed as follows. The $N = 2^n(n \ge 1)$ independent copies
of $W$ are first combined and then split into $N$ bit channels $\{W_N^{(i)}\}_{i=1}^N$ with:
\begin{equation}\label{eq_split}
W_N^{(i)}(y_1^N,u_1^{i-1}|u_i)=\sum_{u_{i+1}^N \in {\mathcal{X}}^{N-i}}\frac{1}{2^{N-1}}W_N(y_1^N|u_1),
\end{equation}
where
\begin{equation}\label{eq_comb}
W_N(y_1^N|u_1^N)=W^N(y_1^N|u_1^NG_N) = \prod_{i=1}^N W(y_i | x_i).
\end{equation}

Mathematically, the encoding is a process to obtain the codeword $x_1^N$ through $x_1^N = u_1^NG$
for given source bits $u_1^N$. The source bits $u_1^N$
consists of the information bits and the frozen bits,
denoted by ${u}_{\mathcal{A}}$ and ${u}_{\bar{\mathcal{A}}}$,
respectively. Frozen bits refer to the fixed transmission
bits which are known to both the transmitter and the receiver.
The set $\mathcal{A}$ includes the indices for the information bits and
$\bar{\mathcal{A}}$ is the complementary set, which can be constructed
as in \cite{arikan_iti09, vardy_it13,mori_icl09,trifonov_itc12,wu_icl14}.
Both sets $\mathcal{A}$ and $\bar{\mathcal{A}}$  are in $\{1, 2,..., N\}$ for polar codes of length $N$.
The source bits $u_1^N$ can be
split as $u_1^N = ({u}_{\mathcal{A}}, {u}_{\bar{\mathcal{A}}})$.
The codeword can then be expressed as
\begin{equation} \label{eq_x2}
\ x_1^N = {u}_{\mathcal{A}}G_{\mathcal{A}}+ {u}_{\bar{\mathcal{A}}}G_{\bar{\mathcal{A}}},
\end{equation}
where $G_{\mathcal{A}}$ is the submatrix of $G_N$ with rows specified by the set $\mathcal{A}$.

\begin{figure}
{\par\centering
{\includegraphics[width=2.5in]{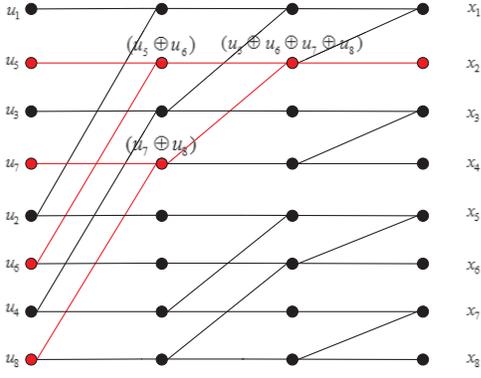}} \par}
\caption{An encoding circuit of the non-systematic polar codes with $N=8$. Signals flow
from the left to the right. Each edge carries a signal of $0$ or $1$.}
\label{fig_encoding_nonsys}
\end{figure}

An encoding diagram is shown in Fig.~\ref{fig_encoding_nonsys}. Each node adds
the signals on all incoming edges from the left and sends the result out on
all edges to the right. The operations are done in the binary field $\mathbb{F}_2$.
One such encoding process is highlighted in Fig.~\ref{fig_encoding_nonsys}
for $x_2=u_5 \oplus u_6 \oplus u_7 \oplus u_8$.
If the nodes in Fig.~\ref{fig_encoding_nonsys} are viewed as memory elements, the encoding
process is to calculate the corresponding binary values to fill all the memory elements from the
left to the right. This view is helpful when it comes to systematic polar codes in the following section.

\subsection{Systematic Polar Codes}\label{sec_background_sys}

The systematic polar code is constructed by
specifying a set of indices of the codeword $x_1^N$ as the indices to convey the information bits.
Denote this set as $\mathcal{B}$ ($|\mathcal{B}| = K$)
and the complementary set as $\bar{\mathcal{B}}$.
The codeword $x_1^N$ is thus split as
$({x}_{\mathcal{B}},{x}_{\bar{\mathcal{B}}})$.
Define a matrix $G_{\mathcal{AB}}$ that is a submatrix of the generator matrix with elements
$\{G_{i,j}\}_{i \in \mathcal{A}, j \in \mathcal{B}}$. 
Splitting $x_1^N$ in  (\ref{eq_x2}) into $(x_{\mathcal{B}}, ~x_{\bar{\mathcal{B}}})$ requires
splitting the matrices $G_{\mathcal{A}}$ and $G_{\bar{\mathcal{A}}}$ as:
\begin{eqnarray}
G_{\mathcal{A}} &=& \left(G_{\mathcal{AB}},~ G_{\mathcal{A}\bar{\mathcal{B}}}\right),\\
G_{\bar{\mathcal{A}}} &=& \left(G_{\bar{\mathcal{A}}\mathcal{B}},~ G_{\bar{\mathcal{A}}\bar{\mathcal{B}}}\right).
\end{eqnarray}
Then $x_1^N$ can be split as the following:
\begin{equation}
\left\{
\begin{aligned} \label{eq_xb_xbc}
{x}_{\mathcal{B}}={u}_{\mathcal{A}}G_{\mathcal{AB}}+{u}_{\bar{\mathcal{A}}}G_{\bar{\mathcal{A}}\mathcal{B}},\\ {x}_{\bar{\mathcal{B}}}={u}_{\mathcal{A}}G_{\mathcal{A\bar{B}}}+{u}_{\bar{\mathcal{A}}}G_{\mathcal{\bar{A}\bar{B}}}.
\end{aligned}
\right.
\end{equation}

We can see from (\ref{eq_xb_xbc}) that, in systematic polar codes, ${x}_{\mathcal{B}}$ plays the role that ${u}_{\mathcal{A}}$ plays in non-systematic polar codes.
Given a non-systematic encoder $(\mathcal{A},u_{\mathcal{\bar{A}}})$,
there exists a systematic encoder $(\mathcal{B},u_{\mathcal{\bar{A}}})$ if $\mathcal{A}$ and $\mathcal{B}$ have the same number of elements and the matrix $G_{\mathcal{AB}}$ is invertible \cite{arikan_icl11}.
Then a systematic encoder can perform the mapping
${x}_{\mathcal{B}} \mapsto {x_1^N}=({x}_{\mathcal{B}},{x}_{\bar{\mathcal{B}}})$. To realize this systematic
mapping, ${x}_{\bar{\mathcal{B}}}$ needs to be computed for any given information bits ${x}_{\mathcal{B}}$. To this
end, we see from (\ref{eq_xb_xbc}) that ${x}_{\bar{\mathcal{B}}}$ can be computed if $u_{\mathcal{A}}$ is known.
The vector $u_{\mathcal{A}}$ can be obtained as the following
\begin{equation} \label{eq_ua}
u_{\mathcal{A}} = (x_{\mathcal{B}}-u_{\bar{\mathcal{A}}}G_{\mathcal{\bar{A}B}})(G_{\mathcal{AB}})^{-1}.
\end{equation}
In \cite{arikan_icl11}, it is shown that $\mathcal{B} = \mathcal{A}$ satisfies all these conditions in order to establish the
one-to-one mapping $x_{\mathcal{B}} \mapsto u_{\mathcal{A}}$. In the rest of the paper, the systematic encoding of polar
codes adopts this selection of $\mathcal{B}$: $\mathcal{B} = \mathcal{A}$. Therefore we can rewrite (\ref{eq_xb_xbc}) as
\begin{equation}
\left\{
\begin{aligned} \label{eq_xb_xbc_2}
{x}_{\mathcal{A}} ={u}_{\mathcal{A}}G_{\mathcal{AA}}+{u}_{\bar{\mathcal{A}}}G_{\bar{\mathcal{A}}\mathcal{A}},\\ {x}_{\bar{\mathcal{A}}} ={u}_{\mathcal{A}}G_{\mathcal{A\bar{A}}}+{u}_{\bar{\mathcal{A}}}G_{\mathcal{\bar{A}\bar{A}}}.
\end{aligned}
\right.
\end{equation}
Note that the submatrix $G_{\mathcal{AA}}$ is a lower triangular matrix with ones at the diagonal.
The entries above the diagonal are all zeros.

Let us go back to the diagram in Fig.~\ref{fig_encoding_nonsys}. For systematic polar codes, the information
bits are now conveyed in the right-hand side in $x_{\mathcal{A}}$. To calculate $x_{\bar{\mathcal{A}}}$,
$u_{\mathcal{A}}$ in the left-hand side needs to be calculated first. Once $u_{\mathcal{A}}$ is obtained,
systematic encoding can be performed in the same way as the non-systematic encoding: performing binary
additions from the left to the right. Therefore, compared with non-systematic encoding, systematic encoding
has an additional round of binary additions from the right to the left. The detailed analysis of systematic
encoding can be found in \cite{li_socc15, hong_icl16}.

\subsection{SC Decoding}
The SC decoding of polar codes follows the same graph as shown in Fig.~\ref{fig_encoding_nonsys}.
The  likelihood ratio (LR) of bit channel $i$ is defined as:
\begin{equation}\label{eq_LR_ui}
L_N^{(i)} = \frac{W_N^{(i)}(y_1^N,u_1^{i-1}|0)}{W_N^{(i)}(y_1^N,u_1^{i-1}|1)}.
\end{equation}
From \cite{arikan_iti09}, it is shown that the transition probability of bit channel $i$ can
be recursively calculated, which results in a recursive calculation of the LRs as:
\begin{equation}\label{eq_LR_odd}
        \begin{split}
           & L_N^{(2i-1)}(y_1^N,\hat{u}_1^{2i-2})= \\
           &\frac{L_{N/2}^{(i)}(y_1^{N/2},\hat{u}_{1,o}^{2i-2}\oplus\hat{u}_{1,e}^{2i-2})
           L_{N/2}^{(i)}(y_{N/2+1}^N,\hat{u}_{1,e}^{2i-2})+1}
           {L_{N/2}^{(i)}(y_1^{N/2},\hat{u}_{1,o}^{2i-2}\oplus\hat{u}_{1,e}^{2i-2})
           +L_{N/2}^{(i)}(y_{N/2+1}^N,\hat{u}_{1,e}^{2i-2})},
        \end{split}
     \end{equation}
\begin{equation}\label{eq_LR_even}
       \begin{split}
          L_N^{(2i)}(y_1^N,\hat{u}_1^{2i-1})=[L_{N/2}^{(i)}(
          & y_1^{N/2},\hat{u}_{1,o}^{2i-2}\oplus\hat{u}_{1,e}^{2i-2})]^{(1-2\hat{u}_{2i-1})}\\
          & \cdot L_{N/2}^{(i)}(y_{N/2+1}^N,\hat{u}_{1,e}^{2i-2}).
       \end{split}
     \end{equation}

\section{Error Correlation Pattern}\label{sec_corre_pattern}
In \cite{arikan_icl11}\cite{li_jsac15}, it is shown that the re-encoding process
of $\hat{x}_1^N={\hat{u}_1^N}G$ after decoding $\hat{u}_1^N$ does not amplify the number of
errors in $\hat{u}_1^N$.
Instead, there are less errors in $\hat{x}_1^N$ than in $\hat{u}_1^N$.
In this section, we state a corollary proven in \cite{li_tvt18} and then provide a proposition to
show the error correlation pattern of the errors in $\hat{u}_1^N$. This pattern is used in Section
\ref{sec_interleaving} to design the CBI scheme.
\newtheorem{corollary}{Corollary}
\newtheorem{lemma}{Lemma}
\begin{corollary}\label{corollary_1}
The matrix $G_{\mathcal{\bar{A}}\mathcal{A}} =\mathbf{0}$.
\end{corollary}
The proof of this corollary can be found in \cite{li_tvt18}.

Now let us define the set $\mathcal{A}_j$ containing the non-zero positions of column $j$ of $G$ as:
\begin{eqnarray}\label{eq_Ai}
\mathcal{A}_j = \left\{ i ~| ~1 \le i \le N \text{~and~} G_{i,j} = 1\right\}.
\end{eqnarray}
Assume the entries of the set $\mathcal{A}_j$ are arranged in the ascending order.
Define $\mathcal{A}_j(a:b)$ as a vector containing element $a$ to element $b$
of the set $\mathcal{A}_j$.
The following lemma can be deduced directly from the construction  and the SC decoding of polar codes.
\begin{lemma}\label{lemma_sc}
Let  $\mathcal{A}_i$ be as defined in (\ref{eq_Ai}) and $j = i-N/2$ ($N/2+1 \le i \le N$).
Then the LR of $\sum_{k \in \mathcal{A}_i}u_{k}$ is directly affected by the decision of
 $\sum_{l \in \mathcal{A}_j(1:N/2)} u_{l}$.
\end{lemma}
\begin{IEEEproof}
To understand the decoding process, let us first look closely at the encoding  process of polar codes.
Fig.~\ref{fig_G_structure} shows the structure of the generator matrix $G = G_N$ and the corresponding
details of the matrix, with respect to the matrix $G_{N/2}$.
Two basic facts of the generator matrix $G_N$ are that:
\begin{itemize}
\item Fact One. Rows $N/2+1$ to $N$ of $G=G_N$ contain two copies of $G_{N/2}$ as: $\left(G_{N/2} ~~ G_{N/2}\right)$.
\item Fact Two. Colums 1 to $N/2$ of $G=G_N$ contain two copies of $G_{N/2}$ as: $\left(\begin{array}{c} G_{N/2} \\  G_{N/2}\end{array}\right)$.
\end{itemize}
In the encoding process, the following two coded bits are achieved:
\begin{eqnarray} \label{eq_xi}
x_i &=& \sum_{k \in \mathcal{A}_i}u_k \\ \label{eq_xj}
x_j &=& \sum_{l \in \mathcal{A}_j(1:N/2)} u_{l} + \sum_{l' \in \mathcal{A}_j(N/2+1:N)} u_{l'}
\end{eqnarray}
Because of Fact One of the generator matrix $G_N$, the  set $\mathcal{A}_j(N/2+1:N)$ ($j=i-N/2$) is the same
as the set $\mathcal{A}_i$.
Therefore the coded bit $x_j$ is:
\begin{equation}
x_j = \sum_{l \in \mathcal{A}_j(1:N/2)} u_{l} + \sum_{k \in \mathcal{A}_i} u_{k}
\end{equation}
The coded bits $x_1^N$ are transmitted over $N$ independent underlying channels $W$, producing corresponding $y_1^N$
observations at the receiver side.
\begin{figure}
\centering
\subfloat[The structure of the generator matrix $G$]{\includegraphics[width=2.0in]{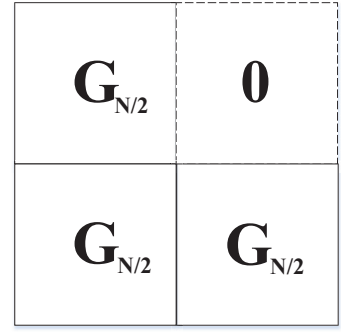}}
\\
\subfloat[Details of $G$]{\includegraphics[width=2.0in]{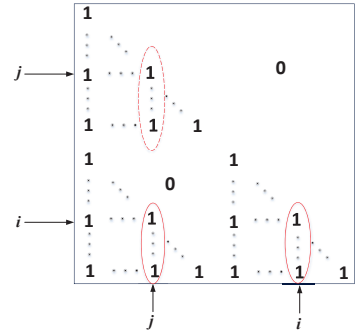}}
\caption{The structure of the genertor matrix $G=G_N = F^{\otimes n}$ and the details of it. The variable
$j$ is spaced by $N/2$ from $i$.}
\label{fig_G_structure}
\end{figure}

In the decoding process, when estimation of ${u}_1^{N/2}$ is done, denoted as $\hat{u}_1^{N/2}$,
then Fact Two can be employed to provide the other $N/2$ observations of the coded bits $x_{N/2+1}^{N}$.
For example, the coded bit $x_i = \sum_{k \in \mathcal{A}_i}u_k$ is observed from the corresponding received
sample $y_i$.

With the estimated $\hat{u}_1^{N/2}$, another observation of $x_i=\sum_{k \in \mathcal{A}_i}u_k$
is readily calculated as: $y_j - \sum_{l \in \mathcal{A}_j(1:N/2)} \hat{u}_{l}$.
This process is captured by the recursive LR calculation in (\ref{eq_LR_even})
where $\sum_{l \in \mathcal{A}_j(1:N/2)} \hat{u}_{l}$ is the estimated decision of the upper left node
and the LR of $\sum_{k \in \mathcal{A}_i} u_{k}$ (at the lower left node) is to be calculated
at that specific connection. Fig.~\ref{fig_recursive_decoding} shows the connection
of that stage.
Therefore, the LR of $\sum_{k \in \mathcal{A}_i} u_{k}$
is affected by the decision of $\sum_{l \in \mathcal{A}_j(1:N/2)} {u}_{l}$ for $j=i-N/2$: if the decision of $\sum_{l \in \mathcal{A}_j(1:N/2)} {u}_{l}$ is incorrect, then the incorrect decision can cause the LR value of $\sum_{k \in \mathcal{A}_i} u_{k}$ incorrect.

\begin{figure}
\centering
{\includegraphics[width=3.0in]{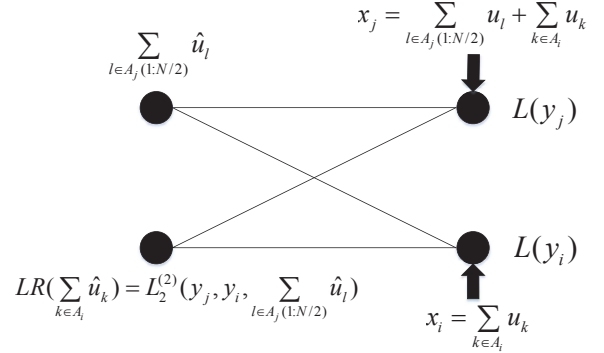}}
\caption{The LR calculation of one stage involving $x_j$ and $x_i$ where $j = i - N/2$.}
\label{fig_recursive_decoding}
\end{figure}
\end{IEEEproof}

\newtheorem{proposition}{Proposition}
\begin{proposition}\label{proposition_1}
Let $\mathcal{A}_i$ be defined as in (\ref{eq_Ai}).
Then, the errors of $\hat{{u}}_{\mathcal{A}_i}$ are dependent (or coupled).
\end{proposition}

Before going into the proof of this Proposition \ref{proposition_1}, we provide an example to explain the meaning of it. As noted in Section \ref{sec_Intro}, the notation $v_{\mathcal{A}}$ is a subvector of $v_1^N$ with elements specified by the set $\mathcal{A}$. Here is an example to show what exactly $\hat{u}_{\mathcal{A}_i}$ is. Let the block length be $N = 16$, the code rate of the polar code be $R = 0.5$, and the underlying channel is the BEC channel
with an erasure probability $0.2$. The set $\mathcal{A}$ is calculated to be $\mathcal{A}=\{8,10,11,12,13,14,15,16\}$. Let $i=10$, then we take the indices of non-zero entries of column $10$ of $G$ as $\mathcal{A}_{10}$, which is a collecting set of indices $10$, $12$, $14$, $16$. Therefore, $\hat{{u}}_{\mathcal{A}_{10}}$ is a subvector of $\hat{u}_1^N$ which contains elements of $\hat{u}_{10}$, $\hat{u}_{12}$, $\hat{u}_{14}$, $\hat{u}_{16}$.

\begin{IEEEproof}\label{proof_1}
We provide proofs of this proposition from two perspectives: 1) From the SC decoding process;
2) From a contradiction perspective with respect to the performance of systematic polar codes.

First, let us prove this proposition from the SC decoding process. The same reasoning in the proof
of Lemma \ref{lemma_sc} 
can be applied here:
the LR of $\sum_{k \in \mathcal{A}_i} u_{k}$ is
directly affected by the decision of $\sum_{l \in \mathcal{A}_j(1:N/2)} {u}_{l}$ for $j=i-N/2$.
With $\mathcal{A}_i = \mathcal{A}_j(N/2+1:N)$, it is exactly saying that the decision of
$\sum_{l \in \mathcal{A}_j(1:N/2)}{u}_{l}$ (from the first half of column $j$) affects the decoding of the $\sum_{l' \in \mathcal{A}_j(N/2+1:N/2)}u_{l'}$ (from another half of column $j$).
Since the recursive LR  calculation of  $u_{j'}$ ($j' \in \mathcal{A}_j(N/2+1:N)$) involves the LR of
 $\sum_{l \in \mathcal{A}_j(1:N/2)}{u}_{l}$ (from the nature of the polar encoding graph),
the decision of bit $u_{j'}$ is therefore affected by the LR of $\sum_{l \in \mathcal{A}_j(1:N/2)}{u}_{l}$.
In other words, any error $u_j$ ($j \in \mathcal{A}_j(1:N/2)$) affects the
decision of the subsequent bit $u_{j'}$ with $j' \in \mathcal{A}_j(N/2+1:N)$.
Therefore the errors in $\hat{u}_{\mathcal{A}_i}$ are correlated.

 Now let us prove the proposition from a contradiction.
 Assume the errors in $\hat{{u}}_{\mathcal{A}_i}$ are independent. For non-systematic polar codes, we define a set $\mathcal{A}_t \subset \mathcal{A}$ containing the indices of the incorrect information bits in an error event. In the same way, we define a set  $\mathcal{A}_{sys,t} \subset \mathcal{A}$ containing the corresponding indices of the information bits in error for systematic polar codes. Let $v_1^N$ be an error indicator vector: a $N$-element vector with $1$s in the positions specified by the error event $\mathcal{A}_t$ and $0$s elsewhere. Let the error probability being:
 $\Pr(v_{m}=1)=p_{m}$. From the independence assumption of errors, it
 is known that $ 0 \le p_m \le 0.5$ for information bits. Correspondingly, we set a vector $q_1^N$ with $1$s in the positions specified by $\mathcal{A}_{sys,t}$ and $0$s elsewhere. From the systematic encoding process, we have $q_1^N={v_1^N}G$. Correspondingly, $q_{\mathcal{A}} = {v_1^N}G(:,\mathcal{A})$ where $G(:,\mathcal{A})$ denotes the submatirx of $G$ composed of the columns specified by $\mathcal{A}$. Since the frozen bits are always correctly determined, $v_{\bar{\mathcal{A}}} = 0_1^{N-K}$ (note that $0_1^{N-K}$ is a zero vector with $N-K$ elements all being zeros). This leads to $q_{\mathcal{A}} = v_{\mathcal{A}}G_{\mathcal{AA}}$. In this way, we convert the errors of non-systematic polar codes and systematic polar codes to the weight of the vectors $v_1^N$ and $q_1^N$.

 Denote the Hamming weight of the vector $v_1^N$ as $w_{H}(v_1^N)$. Specifically, the element $q_i$ ($i \in \mathcal{A}$) is one if $v_{\mathcal{A}_i}$ has an odd number of ones.
With the independent assumption of errors in $\hat{u}_{\mathcal{A}_i}$, the probability that the $i$th information bit $\hat{x}_i$ is in error is
\begin{equation}\label{Pr}
\tilde{p}_i = \frac{1}{2} - \frac{1}{2}\prod_{m=1}^{K_i}(1-2p_m)
\end{equation}\
where $K_i=|\mathcal{A}_i|$. The proof of (\ref{Pr}) is given in Appendix.
In (\ref{Pr}), we can order the probabilities $\{p_m\}_{m=1}^{K_i}$ ($0 \le p_m \le 0.5$) in the ascending order. Applying the Monotone Convergence Theorem to real numbers \cite{yeh_06}, we have:
\begin{flalign}
\begin{split}\label{Pr2}
 \lim_{{K_i}\to\infty}\tilde{p}_i =\lim_{{K_i}\to\infty}[\frac{1}{2}-\frac{1}{2}\prod\limits_{m=1}^{K_i}(1-2p_{m})]=\frac{1}{2}\\
\end{split}
\end{flalign}
Thus, the mean Hamming weight of $q_1^N$: $w_{H}(q_1^N)=\frac{K}{2} \ge w_H(v_1^N)$, meaning the average number of errors of the systematic polar codes is larger than the average number of errors of non-systematic polar codes. This contradicts with the existing results that systematic polar codes
outperform non-systematic polar codes. Thus, we can conclude the errors of $u_{\mathcal{A}_i}$ are dependent.
\end{IEEEproof}
From Proposition \ref{proposition_1}, an error correlation pattern among the errors in $\hat{u}_1^N$ can be
deduced. We call bits $\hat{{u}}_{\mathcal{A}_i}$ the correlated estimated bits. This says that statistically,
the errors of bits $\hat{{u}}_{\mathcal{A}_i}$ are coupled. To show this coupling, we use the same example as the one
after Proposition \ref{proposition_1}.
The number of times the errors of $\hat{u}_{\mathcal{A}_i}$ ($i \in \mathcal{A}$)
happening simultaneously (denoted
by $N_s$)
over the number of times any of the bits $\hat{u}_{\mathcal{A}_i}$ in error (denoted by $N_e$) is called the coupling coefficient, which is equal to $N_s/N_e$.
The coupling coefficients (similar to the correlation coefficient) of bits indicated by non-zero positions
of column $10$, $11$, and $13$
is shown in Table \ref{label_relevant}.  It can be seen from Table I that if there are errors in $\hat{u}_{\mathcal{A}_{10}} = \{ \hat{u}_{10}, \hat{u}_{12}, \hat{u}_{14}, \hat{u}_{16}\}$, then $76\%$ of times these bits errors happen simultaneously, resulting in a coupling coefficient $0.76$ for errors in $\hat{u}_{\mathcal{A}_{10}}$. The coupling coefficients for $\hat{u}_{\mathcal{A}_{11}}$ and $\hat{u}_{\mathcal{A}_{13}}$ are $0.74$ in Table \ref{label_relevant}.

\begin{table}[!t]
\caption{Coupling Effect for $N=16$ and $R=0.5$ in a BEC Channel with an Erasure Probability of $0.2$}
\label{label_relevant}
\centering
\begin{tabular}{|c|c|}
\hline
 Column Index  &  Coupling coefficient \\
\hline
 $10$ & $76$\%  \\
\hline
 $11$ & $74$\%  \\
\hline
 $13$ & $74$\%  \\
\hline
\end{tabular}
\end{table}

To the authors' knowledge, there is no
attempt yet to utilize the error correlation pattern to improve the performance of polar codes. In the next section
of this paper, we propose novel interleaving schemes to break the coupling of errors to improve the BER
performance of polar codes in concatenation systems while still maintaining the low complexity of the SC decoding.

%

\section{The Correlation-Breaking\\
 Interleaving Schemes}\label{sec_interleaving}
In this section we consider interleaving schemes of polar codes (the inner code) with an outer LDPC code
as an example. The introduced schemes  work for all types of outer codes.
From Proposition \ref{proposition_1}, we know
the exact correlated information bits of polar codes.
The interleaving scheme is thus to make sure that the correlated bits of the inner polar codes come from differen LDPC blocks in the transmitter side.
In this way, the de-interleaved LDPC blocks have independent errors.
A blind interleaving (BI) (also known
as direct product) is first introduced, which
breaks all  bits in one LDPC block into different polar code blocks in the interleaver.
Then an improved BI scheme and
a correlation breaking interleaving (CBI) scheme,
that only breaks the correlated bits, are presented.

In this section, we also compare the time complexity and the required memory size of the CBI and the BI schemes.
 The time complexity is in terms of
the decoding delay from the de-interleaving operation: the time from transmitting the first outer code block to
decoding the first outer code block in each round of transmission.

\subsection{The Blind Interleaving Schemes}\label{sec_all_scheme}
In this section, the scheme of scattering all bits in a LDPC block into different polar code blocks is introduced.
The $N_l$ bits of one LDPC block are divided into $N_l$ polar code
blocks, which guarantees that the received error information bits in each LDPC block are independent as they come from different polar code blocks during de-interleaving.

\subsubsection{Direct Product Blind Interleaving}\label{sec_bi_dp}
Denote $c_i^{(j)}$ ($1 \le i \le N_l$, $1 \le j \le K$) as the $i$th coded bit of the $j$th LDPC block. Also
denote $u_k^{(d)}$ as the $k$th information bit of the $d$th polar block. Bits $i$  ($c_i^{(j)}$)  of all
LDPC code blocks form the input vector
to the $i$th polar code encoder. The input bits of the $i$th polar block are arranged
in the order of the LDPC blocks: $u_j^{(i)} = c_i^{(j)}$.
For example, $\{c_1^{(j)}, 1 \le j \le K\}$ of all LDPC blocks produce
the input for the first polar block, and $u_j^{(1)} = c_1^{(j)}$, meaning that
bit one of the $j$th LDPC block is set as the
$j$th input bit of polar block one. This interleaving is called the blind interleaving with direct product (BI-DP).

\begin{figure}
{\par\centering
\resizebox*{3.0in}{!}{\includegraphics[clip]{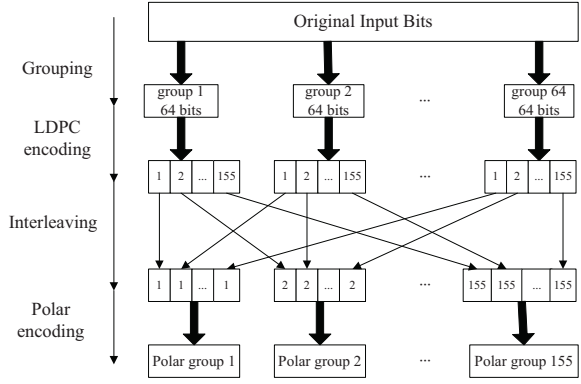}} \par}
\caption{
A blind interleaving scheme with direct product (BI-DP). The block length of the LDPC code is $N_l=155$, and the code rate is $64/155$.
The block length of the polar code is $N=256$, and the code rate is $R=1/4$.
}
\label{fig_all_bits_permutation}
\end{figure}

We give an example in Fig.~\ref{fig_all_bits_permutation} where $K_l=64$ and $N_l=155$.
Polar code in this example has $N=256$, $K=64$ and a code rate $R=1/4$.
Fig.~\ref{fig_all_bits_permutation} is an exact illustration of the BI-DP scheme: bits
one of all LDPC blocks serve as the input to polar block one, bits two of all LDPC blocks serve as the input
to polar block two, and so on.

To compare with the subsequent improved blind interleaving, define a $N_l \times K$ matrix $C$,
that contains the elements of the input of polar blocks. The entry of the
$i$th row and $j$th column is $C_{i,j} = u_{j}^{(i)}$. For the BI-DP scheme, $C_{i,j} = u_j^{(i)} = c_i^{(j)}$.

\subsubsection{Cyclic Direct Product Blind Interleaving}\label{sec_bi_cdp}
One problem with the BI-DP scheme is that for LDPC block $j$, all the coded bits of it are placed as the $j$th
input bits of all polar blocks. For example, all the bits $c_i^{(1)}$ of LDPC block one are the first information
bits of all polar blocks in the receiver side. Given that information bits of polar codes are not equally protected,
it can happen that LDPC block $j$ is exposed to a large amount of errors if bit $j$ of the
polar code is a poorly protected bit in the decoding process. An improved BI, termed cyclic DP (BI-CDP),
is thus introduced below to overcome this problem.

Denote $N_l = n_u K + k_l$, where $n_u$ and $k_l$ are the quotient and the reminder of $N_l$ divided by $K$,
respectively.
Define a basic polynomial $p(x) = j' x^{j'}$ ($0 \le j' \le K-1$). For
the $i$th polar code block ($1 \le i \le n_uK$),  the assignments of the LDPC coded bits to
this polar block can be obtained from the $i'$th ($i' = i-1$) quasi cyclic shift
($0 \le i' \le n_uK-1$):
\begin{equation} \label{eq_bi_cdp_1}
 p^{(i')}(x) = ((j'+ \lfloor i'/K \rfloor K)  x^{(i'+j')}) (\text{mod~}K)
\end{equation}
where `mod' is the modulo operator.
Here the word `quasi' means
that it is not the traditional cyclic shift operation of $x^{i'}p(x)$ because of the jump of the
coefficients every $K$ shifts.

Let $m=(j'+ \lfloor i'/K \rfloor K)+1$, $q=((i'+j') \text{~mod~} K) + 1$ and $l=((m-1) \text{~mod~} K) +1$.
Then the $i'$th cyclic shifted polynomial $p^{(i')}(x)$ carries the $m$th bit of the
$q$th LDPC block $c_{m}^{(q)}$, which is applied to
the $l$th bit of polar block $i=i'+1$, namely $u_{l}^{(i)} = c_{m}^{(q)}$.

This quasi-cyclic
arrangement of LDPC coded bits to the corresponding input bits of polar blocks works for
the first $n_u  K$ polar blocks.
However it  does not work for the last $k_l$ polar blocks because $m = (j'+ n_u K)+1 > N_l$
when $ k_l \le j'  \le K-1 $.

There are many ways to arrange the input for the last $k_l$ polar blocks.
In the following, we propose one possible solution.
Let $i'=i-1=n_u K + i_r$ ($n_uK < i \le N_l$ and $0 \le i_r \le k_l-1$).
For the original polynomial $p(x) = j'x^{j'}$, when $j'=j-1 = i_r$, the
$i'$th cyclic shift is defined as $p^{(i')}(x) = i' x^{j'}$. When $j'=j-1 \neq i_r$,
define a new parameter $j^{''}$ ($0 \le j^{''} \le K-2$) for the other $K-1$
elements of the $i'$th shift of $p(x)$ ($i'=i-1=n_u K + i_r$):
\begin{eqnarray} \nonumber
&& p^{(i')}(x) =\\
&&\left\{  \nonumber
\begin{array}{ll}
i' x^{j'}, &\text{~if~} j' = i_r, \\
(j^{''}\text{~mod~}k_l + n_uK) x^{(i'+j^{''}+1)\text{~mod~}K},  & \text{otherwise}.
\end{array}
\right.\label{eq_bi_cdp_2}
 \\
\end{eqnarray}
It can be verified that the proposed arrangements assign the remaining LDPC coded bits to the last $k_l$ polar blocks.

\begin{figure}
\centering
\subfloat[BI-DP]{\includegraphics[width=2.5in]{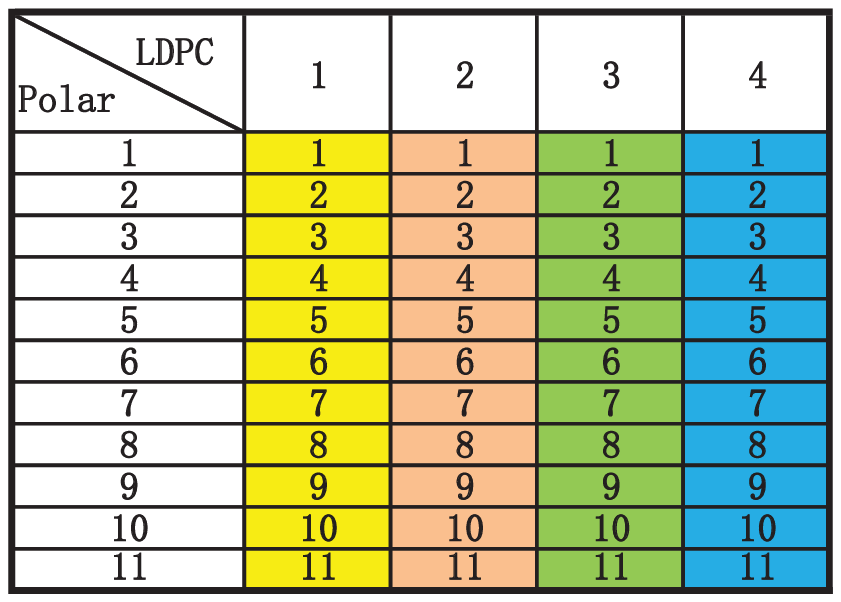}}
\label{fig_bi_dp}
\\
\subfloat[BI-CDP]{\includegraphics[width=2.5in]{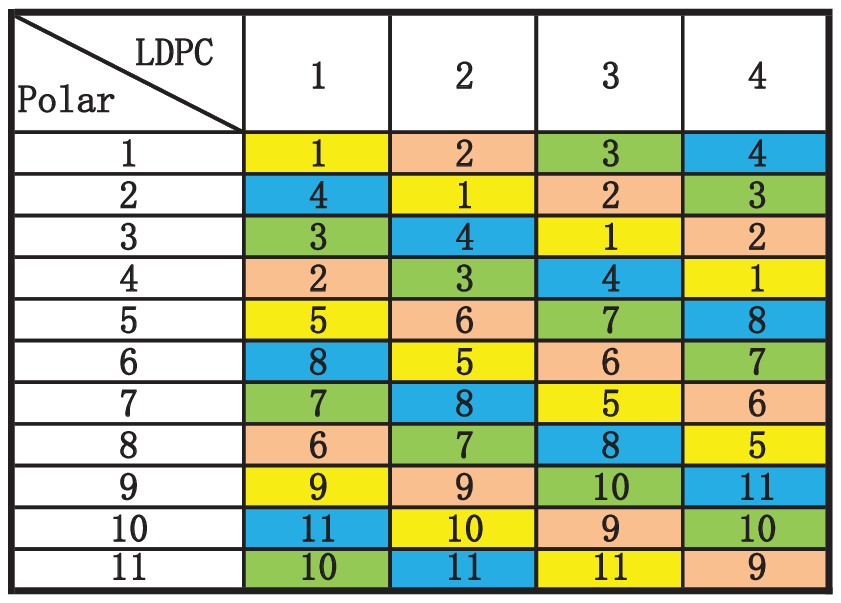}}
\label{fig_bi_cdp}
\caption{An example of the matrix $C$ for $N_l=11$ and $K=4$. The row and column indices are
the indices of polar and LDPC blocks. The entries of one column are the indices of LDPC coded
bits of that specific LDPC block. The four colors of the background corresponding to the four positions of each
polar block.}
\label{fig_bi_dp_cdp}
\end{figure}

This arrangement can be viewed from the matrix $C$ defined  in Sec \ref{sec_bi_dp}.
Fig.~\ref{fig_bi_dp_cdp} shows the assignments of LDPC coded bits to polar blocks, stored
by this matrix $C$.
In this example, $N_l=11$ and $K=4$. For LDPC block $j$ ($1 \le j \le 4$),
the subscript of the coded bits $c_i^{(j)}$ ($1 \le i \le 11$) are stored in column $j$ of the
two tables. For polar block $i$, the input information bits $u_j^{(i)}$ are stored in the
$i$th row of the tables. Since the entries of the tables in Fig.~\ref{fig_bi_dp_cdp}
are the subscripts of $c_i^{(j)}$, the subscripts of the information bits $u_j^{(i)}$
are represented by different colors: yellow is $j=1$ ($u_1^{(i)}$), orange
is $j=2$, green is $j=3$, and blue is $j=4$.

For BI-DP, the assignments of each LDPC coded bits are designed according to Section \ref{sec_bi_dp}. Clearly it can be seen
from the same color of columns of Fig.~\ref{fig_bi_dp_cdp}-(a) that
the coded bits of LDPC block $j$ are assigned to the same bits (the $j$th bits) of all
polar blocks. The assignments of LDPC coded bits for BI-CDP are done according to equations (\ref{eq_bi_cdp_1}) and
(\ref{eq_bi_cdp_2}). Take column 1 (LDPC block 1) of Fig.~\ref{fig_bi_dp_cdp}-(b) as an example.
It is shown that three coded bits (three colored yellow of the column 1) of  LDPC block one are put as the first information bits for three polar blocks (polar block 1, 5, and 9),
two coded bits (two colored orange) are the second information bits of polar blocks 4 and 8, three coded bits (three colored green) are the third information bits of
polar blocks 3, 7 and 11, and three coded
bits (three colored blue) are the fourth information bits of polar blocks 2, 6 and 10.
On the other hand, all eleven coded bits of LDPC block one are the first information bits
of eleven polar blocks for the BI-DP scheme.

Overall, the improved BI-CDP can scatter the LDPC coded bits evenly to the input of polar blocks to
reduce the chance of simultaneous errors. It is expected that the BI-CDP scheme performing better than
the BI-DP scheme.

\subsection{The CBI Scheme}\label{sec_part_scheme}

The two BI schemes in Section \ref{sec_all_scheme} occupies a memory of $[N_l,K]$ received samples.
The  decoding delay of the BI scheme  is $N_l \times N \times T_s$ ($T_s$ is the symbol duration).
From Section \ref{sec_corre_pattern}, we
know that it is not necessary to scatter all bits in a LDPC block into different polar blocks,
since not all bits in a polar block are correlated.
The interleaving scheme in this section is
to make the correlated information bits $u_{\mathcal{A}_i}$ ($1 \le i \le K$) of one polar block come from
different LDPC blocks and the remaining uncorrelated information bits come from one LDPC block in the encoding process. Or in other words, the interleaving
scheme is to scatter only the correlated information bits $u_{\mathcal{A}_i}$ ($1 \le i \le K$) of each polar block
into different LDPC blocks and the uncorrelated information bits of each polar block are scattered into one LDPC block in the receiver side.

The difficulty in designing a CBI scheme is that the sets $\{\mathcal{A}_i\}_{i=1}^K$ are different
for different block lengths and code rates. They are also different for different underlying channels for
which polar codes are designed.
A CBI scheme is dependent on three parameters:
the block length $N$, the code rate $R$, and the underlying channel $W$.
Let us denote a CBI scheme as CBI($N$,$R$,$W$) to show this dependence.
A CBI($N$,$R$,$W$) optimized for one
set of ($N$,$R$,$W$) is not
necessarily optimized for  another set ($N'$,$R'$,$W'$). It may not even work for the set ($N'$,$R'$,$W'$)
if $N'R' \neq NR$. In the following, we provide a CBI scheme which works for any sets of ($N$,$R$,$W$), but
not necessarily optimal for one specific set of ($N$,$R$,$W$).

The set $\mathcal{A}_i$ contains the indices of the non-zero entries of column $i \in \mathcal{A}$.
First, the $K=|\mathcal{A}|$ columns of $G$ are extracted, forming a submatrix $G(:,\mathcal{A})$.
Divide this submatrix further as: $G(:,\mathcal{A}) = [G_{\mathcal{\bar{A}A}} ~~ G_{\mathcal{AA}}]$.
Since the submatrix $G_{\mathcal{\bar{A}A}} = \mathbf{0}$ from Corollary \ref{corollary_1},
it is only necessary to analyze the
submatrix $G_{\mathcal{AA}}$. If a CBI needs to look at each individual set $\mathcal{A}_i$, then
a general CBI is beyond reach. However, we can simplify this problem by dividing
the indices of information bits only into two groups: the correlated bits indices $\mathcal{A}_c$ and the uncorrelated bits indices $\bar{\mathcal{A}_c}$.

Let $\omega_i$ denote the Hamming weight of row $i$ of $G_{\mathcal{AA}}$.
The following proposition can be used to find the sets $\mathcal{A}_c$ and $\bar{\mathcal{A}_c}$.

\begin{proposition}\label{proposition_2}
For the submatrix $G_{\mathcal{AA}}$, define $\mathcal{A}_{cs} = \{i ~| 1 \le i \le K {~\text{and}}~ \omega_i > 1\}$,
and $\bar{\mathcal{A}}_{cs} = \{j ~|~1 \le j \le K {~\text{and}}~ \omega_j = 1\}$.
The corresponding sets of $\mathcal{A}_{cs}$ and $\bar{\mathcal{A}}_{cs}$ with respect to the matrix $G$ are the sets $\mathcal{A}_c$ and $\bar{\mathcal{A}}_c$, respectively.
\end{proposition}

\begin{IEEEproof}
First, let us bear in mind that the submatrix $G_{\mathcal{AA}}$ is a lower triangular matrix as discussed in
Section \ref{sec_background_sys}.
This proposition is equivalent to the following assignment:
\begin{equation} \label{eq_co1}
\begin{cases}
{i \in \bar{\mathcal{A}}_{cs}}, &\text{if $\omega_i=1$}, \\
{i \in \mathcal{A}_{cs}}, &\text{if $\omega_i>1$}.
\end{cases}
\end{equation}

For $\omega_i=1$, there is only one non-zero entry  $G_{i,i} = 1$ for row $i$.
Let $K_c = |\mathcal{A}_{cs}|$ and $K_{uc} = |\bar{\mathcal{A}}_{cs}|$.
Denote the submatrix formed by the rows of $G_{\mathcal{AA}}$ indicated by $\bar{\mathcal{A}}_{cs}$
as $G_{\mathcal{AA}}(\bar{\mathcal{A}}_{cs},:)$.
Then each row of the submatrix $G_{\mathcal{AA}}(\bar{\mathcal{A}}_{cs},:)$
has Hamming weight one. Extract the columns specified by $\bar{\mathcal{A}}_{cs}$ of
$G_{\mathcal{AA}}(\bar{\mathcal{A}}_{cs},:)$ to obtain a matrix denoted as $G_{uc}$. Similar to the process of
extracting $G_{\mathcal{AA}}$ from $G$, the extraction of rows and columns (indicated by $\bar{\mathcal{A}}_{cs}$) from
 $G_{\mathcal{AA}}$ results in a final $K_{uc} \times K_{uc}$ identity
matrix $G_{uc} = I_{K_{uc}}$.

According to Proposition \ref{proposition_1}, the errors in $\hat{u}_{\mathcal{A}_i}$ ($\hat{u}_{\mathcal{A}_c}$)
are coupled. Now that each column of $G_{uc} = I_{K_{uc}}$ has Hamming weight one,
the errors contained in $\hat{u}_{\bar{\mathcal{A}}_{c}}$ are not coupled as indicated by Proposition \ref{proposition_1}.

\end{IEEEproof}

\begin{figure}
{\par\centering
\resizebox*{3.0in}{!}{\includegraphics{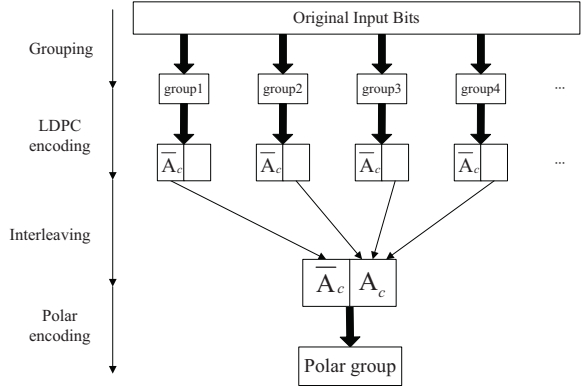}} \par}
\caption{
A general correlation-breaking interleaving scheme.
Here the set $\mathcal{A}_c$ consists of the indices of the correlated bits and the set
$\bar{\mathcal{A}_c}$ is the complementary set of $\mathcal{A}_c$.
}
\label{fig_correlated}
\end{figure}

We use the same  example as before (the one after Proposition \ref{proposition_1})
to show how to use Proposition \ref{proposition_2} to find the sets $\mathcal{A}_c$ and $\bar{\mathcal{A}_c}$.
With Proposition \ref{proposition_2}, we can easily find that $\mathcal{A}_{cs}=\{4,6,7,8\}$ for
the submatrix $G_{\mathcal{AA}}$. Relative to the matrix $G_{16}$, this set is $\mathcal{A}_c=\{12,14,15,16\}$.
The uncorrelated set is thus $\bar{\mathcal{A}_{c}}=\{8,10,11,13\}$.
With the sets $\mathcal{A}_c$ and $\bar{\mathcal{A}_c}$ obtained for any ($N$,$R$,$W$), we can devise a CBI scheme.
Fig.~\ref{fig_correlated} is a general CBI scheme. As in Section \ref{sec_bi_cdp}, let $N_l = n_u K + k_l$ and $K_n=K_c+1$.
For the general CBI scheme, the number of polar blocks, $n_p$,
to transmit $K_n$ LDPC blocks, is expressed as:
\begin{equation}\label{eq_np}
n_p =
\begin{cases}
{(n_u+1) K_n},&\text{if $k_l= 0 ~{\text{or}}~k_l>K_n$}, \\
{n_uK_n+k_l},&\text{otherwise}.
\end{cases}
\end{equation}

The assignment of LDPC coded bits to the polar blocks are similarly done as the BI-CDP scheme,
except that there are coded bits which are put into the uncorrelated positions of the same polar block.
Let $0 \le i' \le n_p - 1$ and $0 \le j' \le K_n-1$.
The general rules to determine the elements of the matrix $C$  are the following:
\begin{itemize}
\item Consider elements of $C$ within the first $n_uK_n$ rows. When $j' = i' {\text{~mod~} K_n }$,  $C_{i,j}$
contains  $K_{uc}$ bits  from LDPC block $j=(j'+1)$. These bits are put into
positions $\bar{\mathcal{A}}_{c}$ of polar block $i=i'+1$.
For the remaining $K_c$ correlated information bits of polar block $i$, it takes coded bits from other different $K_c$ LDPC blocks to put into correlated positions $\mathcal{A}_c$
in the same fashion as the BI-CDP scheme.
\item Consider the rest of the rows (for the remaining $n_p-n_u K_n$ polar blocks). When $j' = i' {\text{~mod~} K_n }$, $C_{i,j}$
contains the remaining bits (smaller than $K_{uc}$) from LDPC block  $j=(j'+1)$.
These bits are also put into positions $\bar{\mathcal{A}}_{c}$ of polar block $i=i'+1$. Polar block $i'$ takes coded
bits from other LDPC blocks for its correlated information bits, similarly to the arrangement of the BI-CDP scheme.
\end{itemize}

Two examples are given in Table \ref{tabel_P1} and Table \ref{tabel_P2} to explain
the assignments for the two cases of (\ref{eq_np}): Table \ref{tabel_P1} is an example of the second case of (\ref{eq_np})
and Table \ref{tabel_P2} is an example of the first case of (\ref{eq_np}).


\begin{table*}
\centering
    \caption{The CBI Scheme for LDPC ($21$,$8$) and Polar ($32$,$16$). The Top Row Contains Indices of LDPC Blocks and the First Column Is the Indices of Polar Blocks. The entries of the table are bit indices of LDPC blocks.}
    \label{tabel_P1}
    \begin{tabular}{|l|l|l|l|l|l|l|l|l|l|l|l|l|l|}
    \hline
    \backslashbox{Polar\kern-1.5em}{\kern-1.5emLDPC}&$1$&$2$&$3$&$4$&$5$&$6$&$7$&$8$&$9$&$10$ \\
    \hline
    $1$&$1:7$&$8$&$9$&$10$&$11$&$12$&$13$&$14$&$15$&$16$\\
    \hline
    $2$&$16$&$1:7$&$8$&$9$&$10$&$11$&$12$&$13$&$14$&$15$\\
    \hline
    $3$&$15$&$16$&$1:7$&$8$&$9$&$10$&$11$&$12$&$13$&$14$\\
    \hline
    $4$&$14$&$15$&$16$&$1:7$&$8$&$9$&$10$&$11$&$12$&$13$ \\
    \hline
    $5$&$13$&$14$&$15$&$16$&$1:7$&$8$&$9$&$10$&$11$&$12$ \\
    \hline
    $6$&$12$&$13$&$14$&$15$&$16$&$1:7$&$8$&$9$&$10$&$11$ \\
    \hline
    $7$&$11$&$12$&$13$&$14$&$15$&$16$&$1:7$&$8$&$9$&$10$ \\
    \hline
    $8$&$10$&$11$&$12$&$13$&$14$&$15$&$16$&$1:7$&$8$&$9$ \\
    \hline
    $9$&$9$&$10$&$11$&$12$&$13$&$14$&$15$&$16$&$1:7$&$8$ \\
    \hline
    $10$&$8$&$9$&$10$&$11$&$12$&$13$&$14$&$15$&$16$&$1:7$ \\
    \hline
    $11$&$17:21$&$17$&$18$&$19$&$20$&$21$&$17$&$18$&$19$&$20$ \\
    \hline
    $12$&$0$&$18:21$&$17$&$18$&$19$&$20$&$21$&$17$&$18$&$19$ \\
    \hline
    $13$&$0$&$0$&$19:21$&$17$&$18$&$19$&$20$&$21$&$17$&$18$ \\
    \hline
    $14$&$0$&$0$&$0$&$20:21$&$17$&$18$&$19$&$20$&$21$&$17$ \\
    \hline
    $15$&$0$&$0$&$0$&$0$&$21:21$&$17$&$18$&$19$&$20$&$21$ \\
    \hline
    \end{tabular}
\end{table*}

\begin{table*}
\centering
\caption{The CBI Scheme for LDPC ($22$,$8$) and Polar ($32$,$8$). The Top Row Contains Indices of LDPC Blocks and the First Column Is the Indices of Polar Blocks. The entries of the table are bit indices of LDPC blocks.}
\label{tabel_P2}
\begin{tabular}{|l|l|l|l|l|l|l|l|l|l|l|l|l|l|}
\hline
\backslashbox{Polar\kern-1.5em}{\kern-1.5emLDPC}&$1$&$2$&$3$&$4$&$5$ \\
\hline
$1$&$1:4$&$5$&$6$&$7$&$8$\\
\hline
$2$&$8$&$1:4$&$5$&$6$&$7$\\
\hline
$3$&$7$&$8$&$1:4$&$5$&$6$\\
\hline
$4$&$6$&$7$&$8$&$1:4$&$5$\\
\hline
$5$&$5$&$6$&$7$&$8$&$1:4$\\
\hline
$6$&$9:12$&$13$&$14$&$15$&$16$\\
\hline
$7$&$16$&$9:12$&$13$&$14$&$15$\\
\hline
$8$&$15$&$16$&$9:12$&$13$&$14$\\
\hline
$9$&$14$&$15$&$16$&$9:12$&$13$\\
\hline
$10$&$13$&$14$&$15$&$16$&$9:12$\\
\hline
$11$&$17:20$&$17$&$18$&$19$&$20$\\
\hline
$12$&$21$&$18:21$&$17$&$18$&$19$\\
\hline
$13$&$22$&$22$&$19:22$&$17$&$18$\\
\hline
$14$&$0$&$0$&$0$&$20:22$&$17$\\
\hline
$15$&$0$&$0$&$0$&$0$&$21:22$\\
\hline
\end{tabular}
\end{table*}

A polar code ($32$,$16$) concatenated with a LDPC code ($21$,$8$) shown in Table \ref{tabel_P1} is the example when $k_l < K_n$.
The correlated set $\mathcal{A}_c = \{16,24,26,27,28,29,30,31,32\}$. Therefore $ K_c = |\mathcal{A}_c| = 9$, $K_n = K_c + 1 = 10$, $n_u = \lfloor N_l/K \rfloor = 1$, and $k_l  = 5 < K_n$.
To transmit $K_n=10$ LDPC blocks, $n_p = n_uK_n + k_l = 15$ polar blocks are required.
In Table \ref{tabel_P1}, the top row contains indices of the LDPC blocks, the first column is the indices of the polar blocks, and the entries of  this table represent the indices of encoded bits of LDPC blocks.
From Table \ref{tabel_P1}, for polar block one, bit $1$ to bit $7$  are taken from LDPC block one,
and the other $9$ bits are bits $8,9,...,16$ from LDPC blocks two to ten, respectively.
The $7$ bits from LDPC block one are placed at the uncorrelated positions $\bar{\mathcal{A}}_c$ of polar block one,
and the other $9$ bits from nine LDPC blocks are arranged at the correlated positions $\mathcal{A}_c$ of polar block one.
The other polar blocks (polar block two to polar block ten) follow the same fashion in collecting the input bits.
These first $n_uK_n$ rows follow the same cyclic assignments of LDPC coded
bits to the inputs of polar blocks as the BI-CDP
scheme. The remaining polar blocks (from polar block eleven to polar block fifteen) collect the remaining bits
of LDPC blocks. For example, although polar block eleven can take $K_{uc} = 7$ uncorrelated bits from LDPC block one,
there are not enough bits left from LDPC block one: only bits $c_{17}$ to $c_{21}$ are left. The assignments of
the correlated positions of polar block eleven follows exactly that of the BI-CDP scheme.

Table \ref{tabel_P2} shows another example when $k_l > K_n$. In this example, the polar code ($32$,$8$) has an $\mathcal{A}_c = \{28,30,31,32\}$ with $K_c =4$ and the LDPC is a ($22$,$8$) code. The parameters are $k_l = 6$ and $K_n = K_c+1 =5$. The total polar blocks $n_p = n_u \times K_n= 3 \times 5$ are used to transmit $K_n = 5$ LDPC blocks.
For both examples, there are $0$s at the left low corner, which means that there are polar positions which are not used. These positions are wasted which are the cost of the universal CBI design.
\subsection{Complexity Analysis}\label{sec_part_delay}

For the CBI scheme, the interleaving requires a  memory to store the decoded LR values
from $n_p$ polar blocks  in order to
do the de-interleaving. The  memory size is therefore $[n_p,K]$.
The decoder needs to wait $n_p$ polar blocks to decode  $K_n$ LDPC blocks. The decoding delay of the outer code  is therefore $n_p \times N \times T_s$, where $T_s$ is the symbol duration in seconds.
For the BI scheme, the memory size is $[N_l,K]$ and the decoding delay is $N_l \times N \times T_s$.

\section{Simulation Results}\label{sec_result}
In this section, simulation results are provided to verify the performance of BI-DP, BI-CDP, and the CBI
scheme. The first example we take is the
same as the BI scheme in Fig.~\ref{fig_all_bits_permutation}.
The LDPC codes used
in this section is the ($155$,$64$)
MacKay code \cite{Mackay_02}, where the code length is  $N_l=155$ and the information bit length is $K_l=64$.
The polar code is ($256$,$64$).
The overall code rate of the LDPC($N_l$,$K_l$)+CBI+POLAR($N$,$K$) concatenation system is $K_l/N_l\times R=0.1$.
The underlying channel is the AWGN channel.
The construction of polar code  is based on \cite{vardy_it13}, which produces the set $\mathcal{A}$.
Then the submatrix $G_{\mathcal{AA}}$ is formed from the generator matrix $G$.
Based on the submatrix   $G_{\mathcal{AA}}$ and Proposition \ref{proposition_2},
for polar code ($256$,$64$), the correlated bits indices is calculated to be $\mathcal{A}_c$ ($K_c = 38$) and the uncorrelated bits indices $\bar{\mathcal{A}_c}$ ($K_{uc} = 26$) are also obtained.
In this example, the occupied memory size of the CBI scheme is $[105,64]$, smaller than $[155,64]$ of the two BI schemes.
The decoding delay of the CBI scheme is $105 \times 256$ symbols,
still smaller than $155 \times 256$ symbols of the BI schemes.

\begin{figure}
{\par\centering
\resizebox*{3.0in}{!}{\includegraphics[clip]{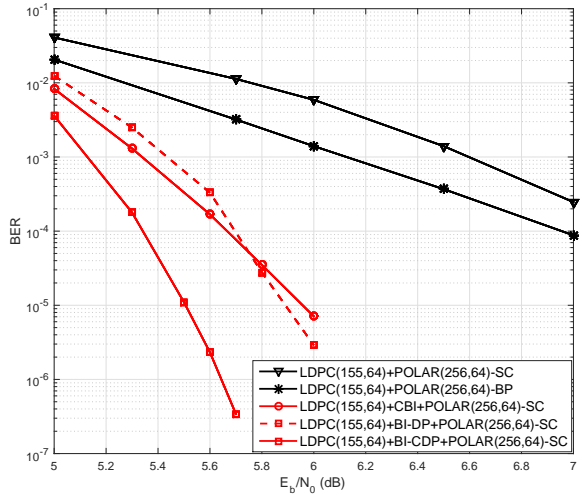}} \par}
\caption{The BER performance of polar code ($256$,$64$) concatenated with a LDPC code in AWGN channels. The LDPC code is the ($155$,$64$) MacKay code.}
\label{fig_simulation_result1}
\end{figure}

The performance of the BI-DP (dashed line with squares) and BI-CDP (solid line with squares)
is shown in Fig.~\ref{fig_simulation_result1}. At a BER $=10^{-5}$, the improved BI-CDP scheme has a $0.4$ dB
advantage over the BI-DP scheme.
To compare with the CBI scheme (the solid  line with circles), two other schemes are also shown in Fig.~\ref{fig_simulation_result1}:
1) the performance of the polar code (SC decoding) directly concatenated with the LDPC code (no interleaving
being performed, denoted by the solid  line with triangles), with a legend of LDPC($155$,$64$)+POLAR($256$,$64$)-SC;
2) the performance of the direct concatenation but with the polar code employing the belief propagation
(BP) decoding (denoted by the solid line with asterisks), with a legend of  LDPC($155$,$64$)+POLAR($256$,$64$)-BP.
At a BER $=10^{-4}$, the LDPC($155$,$64$)+CBI+POLAR($256$,$64$)-SC system achieves $1.4$ dB and $1.2$ dB gains over the direct concatenation systems  LDPC($155$,$64$)+POLAR($256$,$64$)-SC and LDPC($155$,$64$)+POLAR($256$,$64$)-BP, respectively.

Compared with the BI-DP scheme, 
the CBI scheme requires only an additional $0.05$ dB of $E_b/N_0$ to achieve the BER at $10^{-5}$.
Also, the CBI scheme requires a memory size $N_l/n_p = 1.5$ times smaller than that of the BI-DP scheme.
At the same BER level, the BI-CDP scheme outperforms both the BI-DP and the CBI scheme, requiring $0.4$ dB less to achieve
this BER.

\begin{figure}
{\par\centering
\resizebox*{3.0in}{!}{\includegraphics[clip]{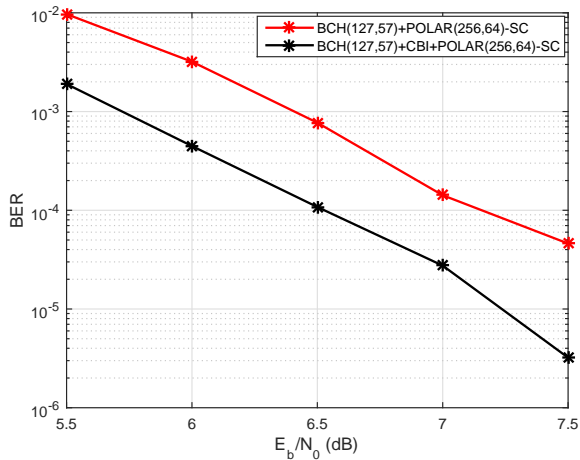}} \par}
\caption{The BER performance of the concatenation scheme in AWGN channels. The BCH code is (127, 57) and the polar code
is (256, 64). }
\label{fig_simulation_result2}
\end{figure}

The proposed CBI scheme can also work with other outer codes, such as BCH codes.
Fig.~\ref{fig_simulation_result2} shows
the result of the polar code ($256$,$64$) with a BCH code ($127$,$57$) where the $127$ and $57$ are the code length and the number of information bits of BCH codes in one code block, respectively.
It can be seen from Fig.~\ref{fig_simulation_result2} that  the CBI scheme employing BCH code as an outer code
has a $0.7$ dB gain over the direct concatenation scheme at a BER $=10^{-4}$.

\section{Conclusion}\label{sec_con}
In this paper, a correlation pattern of bit errors of polar codes with the SC decoding are studied.
Based on the studies, BI-DP, BI-CDP, and CBI schemes are proposed to de-correlate
the coupled bit errors,
while still maintaining the low complexity of the SC decoding of polar codes.
The BI-CDP scheme cyclicly assigns the encoded bits from the outer code to
the input of the inner encoder. As a result, the BI-CDP scheme enjoys a $0.4$ dB
gain over the BI-DP scheme for the presented results in the paper.
The proposed novel CBI scheme  has a much better performance than the direct concatenation schemes.
Compared with the BI-DP scheme, the CBI scheme
also achieves a comparable BER performance while requiring a smaller memory size and a shorter
decoding delay.
Simulation results verify the theories and the proposed schemes in the paper.

\appendix [Proof of Equation(\ref{Pr})] \label{Pr_eq10}
\begin{IEEEproof}
Given a sequence of $M$ independent binary digits $v_1^M$ where the probability $\Pr(v_m=1)=p_m$, then the probability that $v_1^M$ contains an odd number of $1$'s (denoted by $P_M$) is
\begin{equation}\label{Pr_l}
P_M=\frac{1}{2} - \frac{1}{2}\prod_{m=1}^{M}(1-2p_m).
\end{equation}\

We use induction to prove it.
First, let $M=1$, then $P_1=p_m=\frac{1}{2} - \frac{1}{2}\prod_{m=1}^{1}(1-2p_m)$.
Next assume when $M=k_m$, (\ref{Pr_l}) holds. That is:
$P_{k_m}=\frac{1}{2} - \frac{1}{2}\prod_{m=1}^{k_m}(1-2p_m)$.
 Now let us prove that when $M=k_m+1$, (\ref{Pr_l}) still holds:
\begin{equation}\label{Pr_km1}
P_{k_m+1}=\frac{1}{2} - \frac{1}{2}\prod_{m=1}^{k_m+1}(1-2p_m).
\end{equation}
Starting from $P_{k_m}$, $P_{k_m+1}$ can be
derived as the following:
\begin{align*}
P_{k_m+1}&=p_{k_m+1} \times (1-P_{k_m})+P_{k_m} \times (1-p_{k_m+1}) \\
&        =p_{k_m+1}-2 \times P_{k_m} \times p_{k_m+1}+P_{k_m} \\
&        =p_{k_m+1} \times \prod_{m=1}^{k_m}(1-2p_m)+\frac{1}{2} - \frac{1}{2}\prod_{m=1}^{k_m}(1-2p_m).\\
\end{align*}
Let us extend the right-hand side of (\ref{Pr_km1}) as the following:
\begin{align*}
&\frac{1}{2} - \frac{1}{2}\prod_{m=1}^{k_m+1}(1-2p_m)\\
&                    =\frac{1}{2} - \frac{1}{2}\prod_{m=1}^{k_m}(1-2p_m)\times(1-2p_{k_m+1})\\
&                    =p_{k_m+1} \times \prod_{m=1}^{k_m}(1-2p_m)+\frac{1}{2} - \frac{1}{2}\prod_{m=1}^{k_m}(1-2p_m).\\
\end{align*}
which is equal to the one derived from $P_{k_m}$.
Therefore, equation (\ref{Pr}) is proven from the induction.
\end{IEEEproof}

\ifCLASSOPTIONcaptionsoff
  \newpage
\fi

\footnotesize
\bibliographystyle{IEEEtran}
\bibliography{ref_polar_tc18}

\begin{thebibliography}{10}
\providecommand{\url}[1]{#1}
\csname url@samestyle\endcsname
\providecommand{\newblock}{\relax}
\providecommand{\bibinfo}[2]{#2}
\providecommand{\BIBentrySTDinterwordspacing}{\spaceskip=0pt\relax}
\providecommand{\BIBentryALTinterwordstretchfactor}{4}
\providecommand{\BIBentryALTinterwordspacing}{\spaceskip=\fontdimen2\font plus
\BIBentryALTinterwordstretchfactor\fontdimen3\font minus
  \fontdimen4\font\relax}
\providecommand{\BIBforeignlanguage}[2]{{%
\expandafter\ifx\csname l@#1\endcsname\relax
\typeout{** WARNING: IEEEtran.bst: No hyphenation pattern has been}%
\typeout{** loaded for the language `#1'. Using the pattern for}%
\typeout{** the default language instead.}%
\else
\language=\csname l@#1\endcsname
\fi
#2}}
\providecommand{\BIBdecl}{\relax}
\BIBdecl

\bibitem{arikan_iti09}
E.~Ar$\i$kan, ``{Channel polarization: A method for constructing
  capacity-achieving codes for symmetric binary-input memoryless channels},''
  \emph{IEEE Trans. Inf. Theory}, vol.~55, no.~7, pp. 3051--3073, Jul. 2009.

\bibitem{arikan_icl11}
------, ``{Systematic polar coding},'' \emph{IEEE Commun. Lett.}, vol.~15,
  no.~8, pp. 860--862, Aug. 2011.

\bibitem{vardy_it13}
I.~Tal and A.~Vardy, ``{How to construct polar codes},'' \emph{IEEE Trans. Inf.
  Theory}, vol.~59, no.~10, pp. 6562--6582, Oct. 2013.

\bibitem{mori_icl09}
R.~Mori and T.~Tanaka, ``{Performance of polar codes with the construction
  using density evolution},'' \emph{IEEE Commun. Lett.}, vol.~13, no.~7, pp.
  519--521, Jul. 2009.

\bibitem{trifonov_itc12}
P.~Trifonov, ``{Efficient design and decoding of polar codes},'' \emph{IEEE
  Trans. Commun.}, vol.~60, no.~11, pp. 3221--3227, Nov. 2012.

\bibitem{wu_icl14}
D.~Wu, Y.~Li, and Y.~Sun, ``{Construction and block error rate analysis of
  polar codes over AWGN channel based on Gaussian approximation},'' \emph{IEEE
  Commun. Lett.}, vol.~18, no.~7, pp. 1099--1102, Jul. 2014.

\bibitem{C.Z_itsp13}
C.~Zhang and K.~K. Parhi, ``{Low-latency sequential and overlapped
  architectures for successive cancellation polar decoder},'' \emph{IEEE Trans.
  Signal Process.}, vol.~61, pp. 2429--2441, May 2013.

\bibitem{vardy_jsac14}
G.~Sarkis, P.~Giard, A.~Vardy, C.~Thibeault, and W.~Gross, ``{Fast polar
  decoders: Algorithm and implementation},'' \emph{IEEE J. Sel. Areas Commun.},
  vol.~32, no.~5, pp. 946--957, May 2014.

\bibitem{C.Z_itcs14}
C.~Zhang and K.~K. Parhi, ``{Latency analysis and architecture design of
  simplified SC polar decoders},'' \emph{IEEE Trans. Circuits Syst. II, Exp.
  Briefs}, vol.~61, pp. 115--119, Feb. 2014.

\bibitem{arikan_icl08}
E.~Ar$\i$kan, ``{A performance comparison of polar codes and reed-muller
  codes},'' \emph{IEEE Commun. Lett.}, vol.~12, no.~6, pp. 447--449, Jun. 2008.

\bibitem{urbanke_isit09}
N.~Hussami, S.~Korada, and R.~Urbanke, ``{Performance of polar codes for
  channel and source coding},'' in \emph{Proc. IEEE Int. Symp. Inform. Theory
  (ISIT)}, Jun. 2009, pp. 1488--1492.

\bibitem{vardy_it15}
I.~Tal and A.~Vardy, ``{List decoding of polar codes},'' \emph{IEEE Trans. Inf.
  Theory}, vol.~61, no.~5, pp. 2213--2226, May 2015.

\bibitem{niu_itc13}
K.~Chen, K.~Niu, and J.~Lin, ``{Improved successive cancellation decoding of
  polar codes},'' \emph{IEEE Trans. Commun.}, vol.~61, no.~8, pp. 3100--3107,
  Aug. 2013.

\bibitem{eslami_isit11}
A.~Eslami and H.~Pishro-Nik, ``{A practical approach to polar codes},'' in
  \emph{Proc. IEEE Int. Symp. Inform. Theory (ISIT)}, Jul. 2011, pp. 16--20.

\bibitem{guo_isit14}
J.~Guo, M.~Qin, A.~G. i~Fabregas, and P.~H. Siegel, ``{Enhanced belief
  propagation decoding of polar codes through concatenation},'' in \emph{Proc.
  IEEE Int. Symp. Inform. Theory (ISIT)}, Jun. 2014, pp. 2987 -- 2991.

\bibitem{effros_isit10}
M.~Bakshi, S.~Jaggi, and M.~Effros, ``Concatenated polar codes,'' in
  \emph{Proc. IEEE Int. Symp. Inform. Theory (ISIT)}, Jun. 2010, pp. 918--922.

\bibitem{jiang_cl16}
T.~Wang, D.~Qu, and T.~Jiang, ``Parity-check-concatenated polar codes,''
  \emph{IEEE Commun. Lett.}, vol.~20, no.~12, pp. 2342--2345, Dec. 2016.

\bibitem{song_el17}
J.~Park, I.~Kim, and H.~Song, ``Construction of parity-check-concatenated polar
  codes based on minimum hamming weight codewords,'' \emph{Electron. Lett.},
  vol.~53, no.~14, pp. 924--926, Jul. 2017.

\bibitem{Stefan_infoTheory02}
R.~J. Stefan~H\"ost and V.~V. Zyablov, ``Woven convolutional codes i: Encoder
  properties,'' vol.~48, no.~1, pp. 149--161, Jan. 2002.

\bibitem{lin_ecc04}
S.~Lin and D.~J. Costello, \emph{Error Control Coding}, 2nd~ed.\hskip 1em plus
  0.5em minus 0.4em\relax Pearson Prentice Hall, 2004.

\bibitem{ya_vtc16}
Y.~Meng, L.~Li, and Y.~Hu, ``{A novel interleaving scheme for polar codes},''
  in \emph{Proc. IEEE Veh. Technol. Conf. Fall (VTC-Fall)}, Sep. 2016, pp.
  1--5.

\bibitem{li_socc15}
L.~Li and W.~Zhang, ``{On the encoding complexity of systematic polar codes},''
  in \emph{Proc. IEEE Int. Syst.-on-Chip Conf. (SOCC)}, Sep. 2015, pp.
  508--513.

\bibitem{hong_icl16}
H.~Vangala, Y.~Hong, and E.~Viterbo, ``{Efficient algorithms for systematic
  polar encoding},'' \emph{IEEE Commun. Lett.}, vol.~20, no.~1, pp. 17--20,
  Jan. 2016.

\bibitem{li_jsac15}
L.~Li, W.~Zhang, and Y.~Hu, ``{On the error performance of systematic polar
  codes},'' [Online]. Available: \url{http://arxiv.org/abs/1504.04133}, 2015.

\bibitem{li_tvt18}
L.~Li, Z.~Xu, and Y.~Hu, ``Channel estimation with systematic polar codes,''
  \emph{IEEE Trans. Veh. Technol.}, vol.~67, no.~6, pp. 4880--4889, Jun. 2018.

\bibitem{yeh_06}
J.~Yeh, \emph{{Real analysis: Theory of measure and integration}},
  2nd~ed.\hskip 1em plus 0.5em minus 0.4em\relax World Scientific Publishing
  Co., 2006.

\bibitem{Mackay_02}
D.~J. Mackay, ``{Good error-correcting codes based on very sparse matrices},''
  [Online]. Available:
  \url{http://www.inference.phy.cam.ac.uk/mackay/CodesGallager.html}, Jul.
  2002.

\end{thebibliography}

\begin{IEEEbiography}
[{\includegraphics[width=1in,height=1.25in,clip,keepaspectratio]{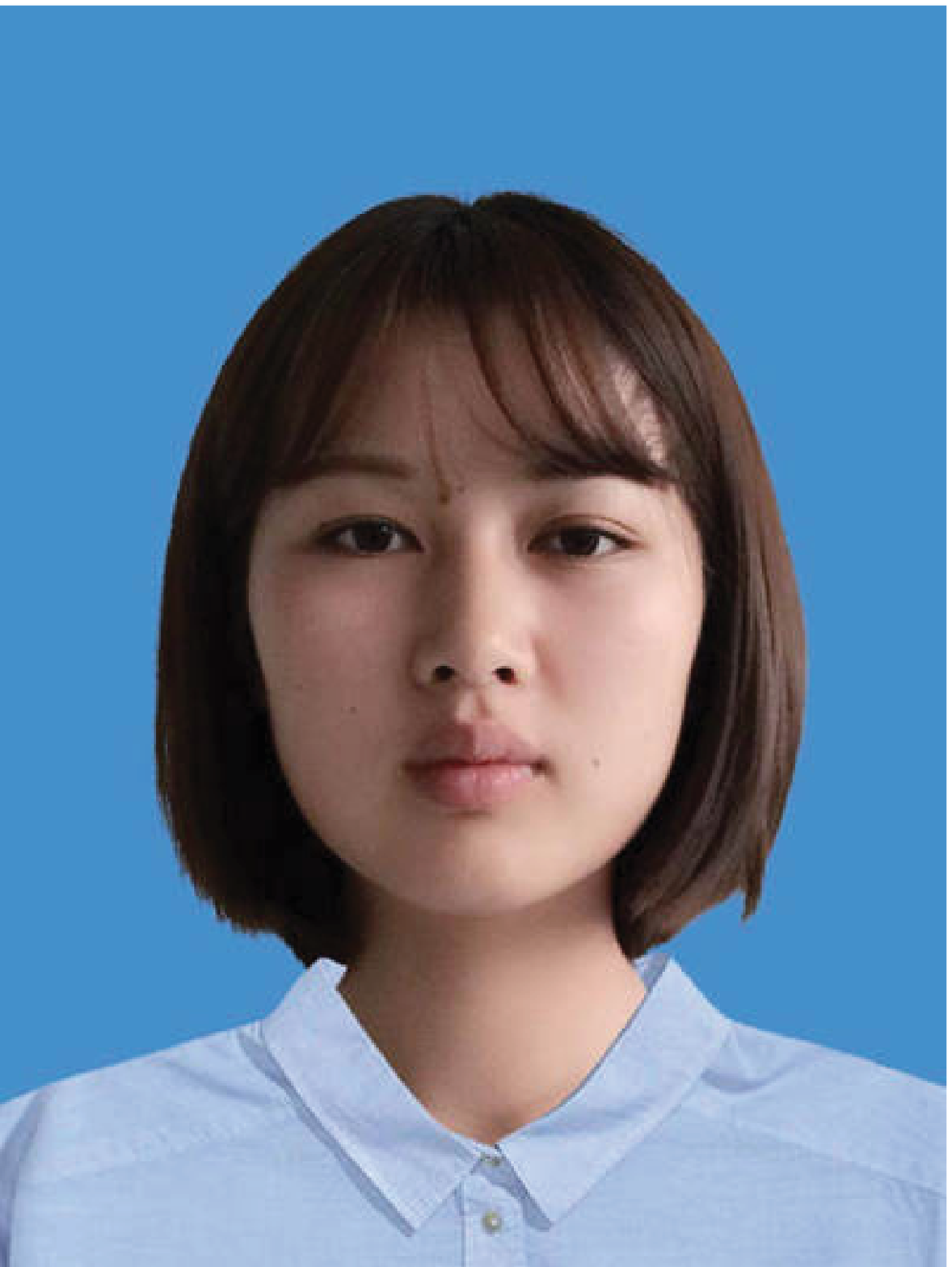}}]{Ya Meng}
(S'16) is currently pursuing the B.S.
degree in the School of Electronics and Information Engineering, Anhui University. Her research interest is in polar codes. Specifically, her research is to find the error propagation pattern of the SC decoding of polar codes, and to use this error pattern to improve the performance of the SC decoding. She was also awarded the Third Place of the Ninth International Students' Innovation and Entrepreneurship Competition (I CAN).
\end{IEEEbiography}
\begin{IEEEbiography}
[{\includegraphics[width=1in,height=1.25in,clip,keepaspectratio]{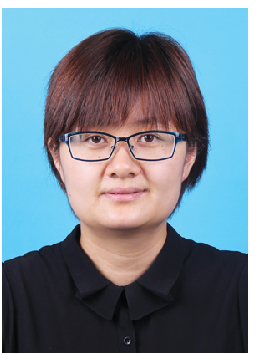}}]{Liping Li}
(S'07-M'15) is now an associate professor of the Key Laboratory of Intelligent Computing and Signal Processing of the Ministry of Education of China, Anhui University. She got her PhD in Dept. of Electrical and Computer Engineering at North Carolina State University, Raleigh, NC, USA, in 2009. Her current research interest is in channel coding, especially polar codes. Dr. Li's research topic during her PhD studies was multiple-access interference analysis and synchronization for ultra-wideband communications. Then she worked on a LTE indoor channel sounding and modeling project in University of Colorado at Boulder, collaborating with Verizon. From 2010 to 2013, she worked at Maxlinear Inc. as a staff engineer in the communication group. At Maxlinear, she worked on SoC designs for the ISDB-T standard and the DVB-S standard, covering modules on OFDM and LDPC. In Sept. 2013, she joined Anhui University and started her research on polar codes until now.
\end{IEEEbiography}
\begin{IEEEbiography}
[{\includegraphics[width=1in,height=1.25in,clip,keepaspectratio]{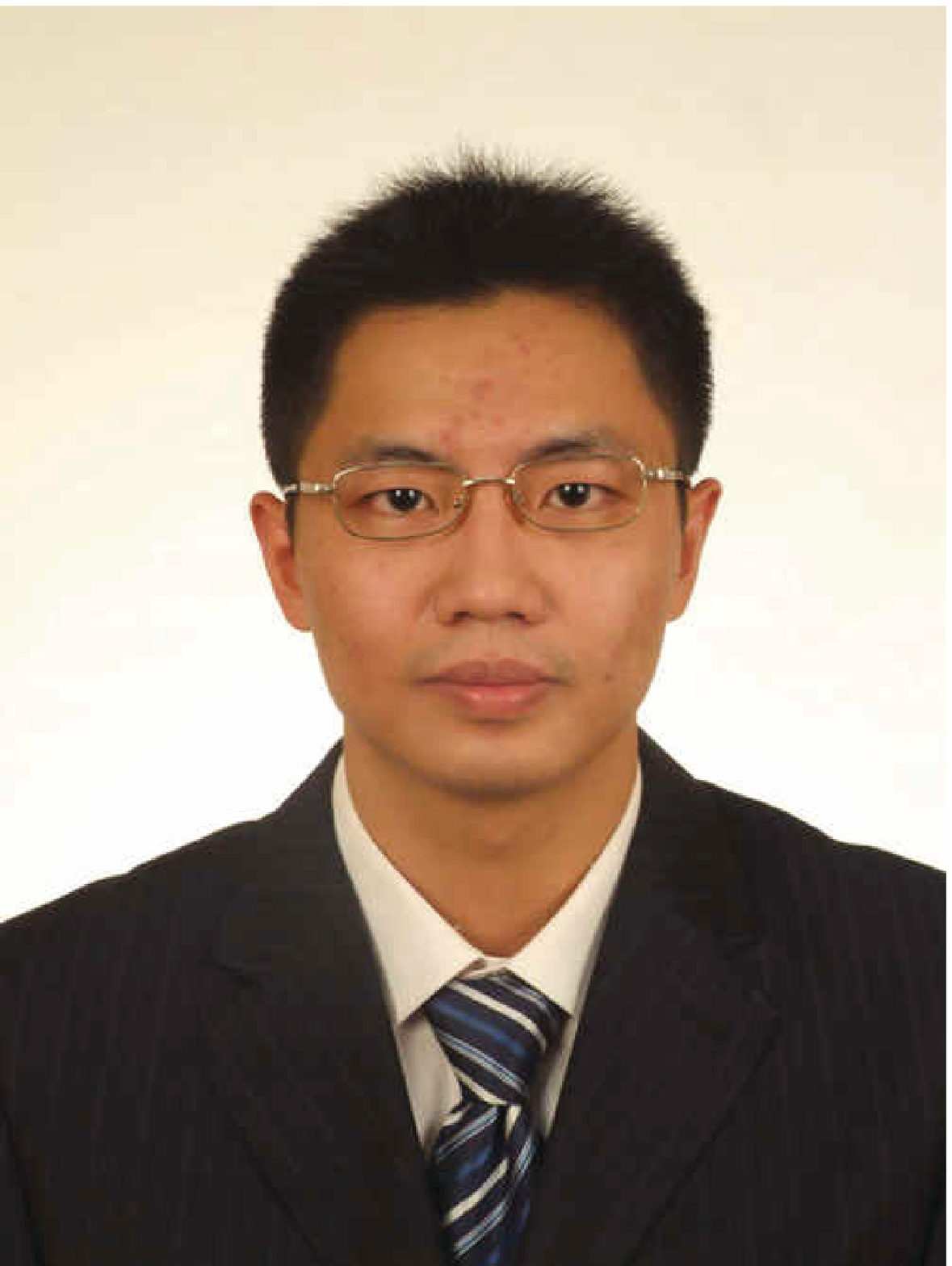}}]{Chuan Zhang}
(S'07-M'13) received the B.E.degree (summa cum laude) in microelectronics and the M.E. degree in very-large scale integration (VLSI) design from Nanjing University, Nanjing, China, in 2006 and 2009, respectively, and the M.S.E.E. and Ph.D. degrees from the Department of Electrical and Computer Engineering, University of Minnesota, Twin Cities (UMN), USA, in 2012. He is currently an Associate Professor with the National Mobile Communications Research Laboratory, School of Information Science and Engineering, Southeast University, Nanjing. His current research interests include low-power high-speed VLSI design for digital signal processing and digital communication, bio-chemical computation and neuromorphic engineering, and quantum communication. He is a member of the Seasonal School of Signal Processing and the Design and Implementation of Signal Processing Systems, the TC of the IEEE Signal Processing Society, and Circuits and Systems for Communications, the VLSI Systems and Applications, the Digital Signal Processing, and the IEEE Circuits and Systems Society. He was a co-recipient of the Best Paper Award of the IEEE Asia Pacific Conference on Circuits and Systems (APCCAS) in 2016, the Best (Student) Paper Award of the IEEE International Conference on DSP in 2016, three excellent paper awards and the Excellent Poster Presentation Award of the International Collaboration Symposium on Information Production and Systems in 2016 and 2017, two Best (Student) Paper Award at the IEEE International Conference on ASIC in 2015 and 2017, respectively, the Best Paper Award Nomination of the IEEE Workshop on Signal Processing Systems in 2015, the Merit (Student) Paper Award of the IEEE APCCAS in 2008. He received the Three-Year University-Wide Graduate School Fellowship of UMN and the Doctoral Dissertation Fellowship of UMN.
\end{IEEEbiography}
\clearpage

\end{document}